\documentclass[
    ,final            
  ]
  {aipproc}

\layoutstyle{6x9}

\usepackage{float}
\usepackage{epsfig}
\usepackage{subfigure}

\newcommand{\cD}{{\cal D}}

\newcommand{\no}{\noindent}
\newcommand{\be}{\begin{eqnarray}}
\newcommand{\ee}{\end{eqnarray}}
\newcommand{\hk}{\hspace{0.1cm}}

\newcommand{\rk}{\right)}
\newcommand{\lk}{\left(}

\begin{document}

\title{Hamiltonian approach to Yang-Mills theory in Coulomb gauge\thanks{Invited
      plenary talk given by H. Reinhardt
       at the conference ``Quark Confinement and the
      hadron spectrum VII'', Ponta del Gada, Portugal, 2.-7.9.2006.}}


\author{H. Reinhardt, D. Epple, W. Schleifenbaum}{
  address={Universit\"at T\"ubingen, Institut f\"ur Theoretische Physik, Auf der
  Morgenstelle 14, 72076 T\"ubingen, Germany}
}



\begin{abstract}
Recent results obtained within the Hamiltonian approach to continuum Yang-Mills
theory in Coulomb gauge are reviewed. 
 \end{abstract}

\maketitle


\section{Introduction}


\no
There have been many attempts in the past to solve the Yang-Mills Schr\"odinger
equation using gauge invariant wave functionals. Unfortunately, all these
approaches have not been blessed with much success\footnote{Recent progress has
been made, however, in $D = 2 + 1$, see ref.\ \cite{Leigh:2005dg}.}. The reason is that it is extremely
difficult to work with gauge invariant wave functionals. A much more economic
way is to explicitly resolve Gauss' law (which ensures gauge invariance of the
wave functional) by fixing the gauge. For this purpose, the Coulomb gauge
  $\vec{\partial}\cdot \vec{A} = 0$ is particularly convenient. In my talk, I would like to present a variational
solution of the Yang-Mills Schr\"odinger equation in Coulomb gauge \cite{Feuchter:2004mk}. Let me start
by briefly summarizing the essential ingredients of the quantization of
Yang-Mills theory in Coulomb gauge.

\no
In Coulomb gauge the
  space of (transversal) gauge orbits has  a non-trivial metric, 
  which is given by the Faddeev-Popov
  determinant  $J (A) = Det (- \hat{D}_i
\partial_i )$, where $\hat{D}^{a b}_i = \delta^{a b} \partial_i  + \hat{A}^{a
b}_i \hk , \hk \hat{A}^{a
b}_i = f^{a c b} A^c_i$ denotes the covariant derivative in the adjoint
representation of the gauge group ($f^{a c b}$ is the structure constant). In
Coulomb gauge the Yang-Mills Hamiltonian is given by \cite{Christ:1980ku}
\be
\label{G1}
H  =  \frac{1}{2} \int J^{- 1}  \Pi J  \Pi + \frac{1}{2} \int B ^2
 + \frac{g^2}{2} \int J^{-1}\rho (- \hat{D} \partial)^{- 1} (- \partial^2) (-
\hat{D} \partial)^{- 1} J\rho \hk ,
\ee
where $\Pi^a_i (x) = \delta / i \delta A^a_i (x)$ denotes the momentum operator,
representing the color electric field, $B$ is the color magnetic field and $\rho^a (x) = - \hat{A}^{a b}_i (x)
\Pi^b_i (x)$ is the non-Abelian color charge of the gauge field. The first term
in the Hamiltonian is the Laplacian in curved space, the second term represents
the potential and the last term arises from the longitudinal momentum part of
the kinetic energy after resolving Gauss' law. This term is usually referred to
as the Coulomb term, since in the Abelian case it reduces to the ordinary Coulomb
potential.

\no
  
\section{Variational solution of the Yang-Mills Schr\"odinger equation}



\no
We have performed a variational solution of the Yang-Mills Schr\"odinger
equation for the vacuum using the following ansatz for the wave functional \cite{Feuchter:2004mk}
\be
\label{G2}
\Psi [A] = {\cal N} J [A]^{- \alpha} \exp \lk - \frac{1}{2} \int A \omega A \rk
\hk 
\ee
with\footnote{This ansatz is motivated by the form of the wave function of a
 point particle in a spherical symmetric potential in a zero angular momentum
 (s-)state $\Psi (r) = \frac{u (r)}{r}$,
 where $J (r) = r^2$. In ref. \cite{Szczepaniak:2001rg} the ansatz (\ref{G2})  with $\alpha = 0$ was used.}
  $\alpha = \frac{1}{2}$, where $\omega ( | x - x' |)$ is 
 the variational
kernel determined by minimizing the energy
\be
\label{G3}
\langle \Psi | H | \Psi \rangle = \int \cD A\: J [A] \Psi^* [A]\: H\: \Psi [A] \hk .
\ee
This gives rise to a set of coupled Dyson-Schwinger equations for the ghost
propagator
\be
\label{G4}
G^{a b} (x, x') = \langle \Psi | \langle x, a | (- \hat{D} \partial)^{- 1} | x',
b \rangle | \Psi \rangle = \frac{1}{g} (- \partial^2)^{-1}d(x,x') \hk 
\ee
with $d$ being the ghost form factor, and the gluon propagator 
 \be
 \label{G5}
 D^{a b}_{i j} (x, x') = \langle \Psi | A^a_i (x) A^b_j (x) | \Psi \rangle =
 \frac{1}{2} \delta^{a b} t_{i j} (x) \omega^{- 1} (x, x') \hk ,\, \, \hk t_{i j} (x)
 = \delta_{i j} - \frac{\partial^x_i \partial^x_j}{\partial^2_x} \hk .
 \ee
  The gluon Dyson-Schwinger equation has the form of the dispersion relation of a
 relativistic particle
 \be
 \label{G6}
 \omega^2 (k) = k^2 + \Sigma^2 (k) \hk ,
 \ee
 where the gluon self-energy $\Sigma$ is dominated by the ghost loop
 \be
 \label{G7}
 \chi^{a b}_{i j} (x, x') = - \frac{1}{2} \left\langle \Psi |\right. \frac{\delta^2 \ln
 J [A]}{\delta A^a_i (x) \delta A^b_j (y)} \left.| \Psi \right\rangle \hk ,
 \ee
 which is a measure for the curvature of the space of transversal gauge orbits
 and which in the following will be referred to as curvature. Due to the ansatz for the ghost form factor $d$ the coupling constant $g$ drops 
 out from the Dyson-Schwinger equations which is a specific feature of the one-loop approximation.
 
 \no
In ref. \cite{Schleifenbaum:2006bq} an infrared analysis of the Dyson-Schwinger equations was
performed without resorting to the angular approximation. Using the infrared
ans\"atze
\be
\label{1}
D (p) = \frac{1}{p^{2 + \alpha}} \hk , \hk G (p) = \frac{B}{p^{2 + \beta}}
\ee
and implementing the horizon condition $d^{- 1} (0) = 0$, first given in ref.\ \cite{Zwanziger:1995cv}, one finds from the
ghost Dyson-Schwinger equation the sum rule
\be
\label{2}
\alpha + 2 \beta = d - 4 \hk ,
\ee
where $d$ is the number of spatial dimensions, see also ref.\ \cite{Zwa02}. The sum rule (\ref{2}) is due to the non-renormalization of the ghost-gluon vertex which is a feature of both the Landau gauge \cite{Tay71} and the Coulomb gauge \cite{Fischer:2005qe,Schleifenbaum:2006bq}.
Inserting this relation into the gluon Dyson-Schwinger equation, one obtains 
for the infrared
exponent a unique solution $\beta = 0.4$ in $d = 2$  and two solutions in $d = 3$,
\be
\label{3}
\beta_1 = 0.796 \hk , \hk \beta_2 = 1.0 \hk .
\ee
Only $\beta_2$ was previously found in the angular
approximation \cite{Feuchter:2004mk}, while $\beta_1$ corresponds to the numerical
solutions reported in ref. \cite{Feuchter:2004mk}, where $\beta = 0.85$ was found. Recently
the full (numerical) solution corresponding to the infrared exponent $\beta_2$
was also found \cite{Epple} and is shown in fig.\ 1.

\begin{figure}[ht]
\label{fig1}
  \includegraphics[scale=0.57]{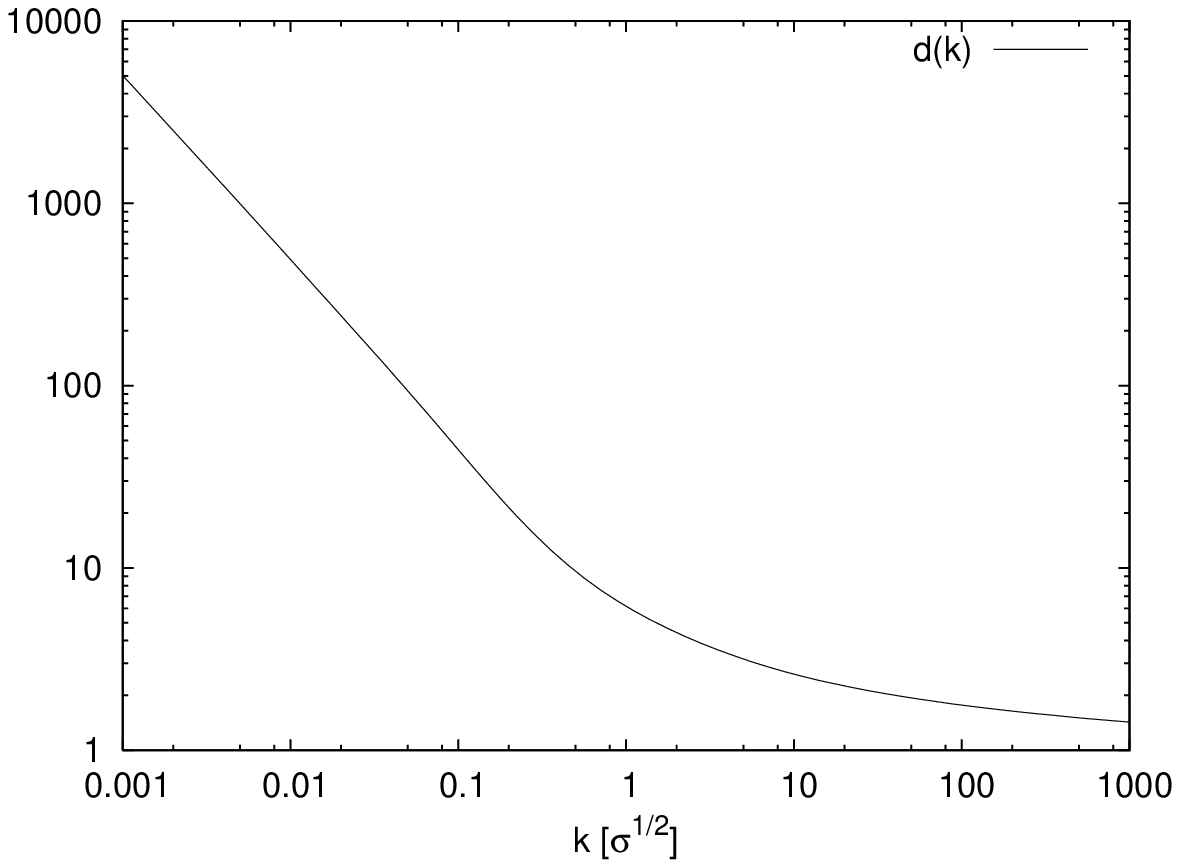}
  \includegraphics[scale=0.57]{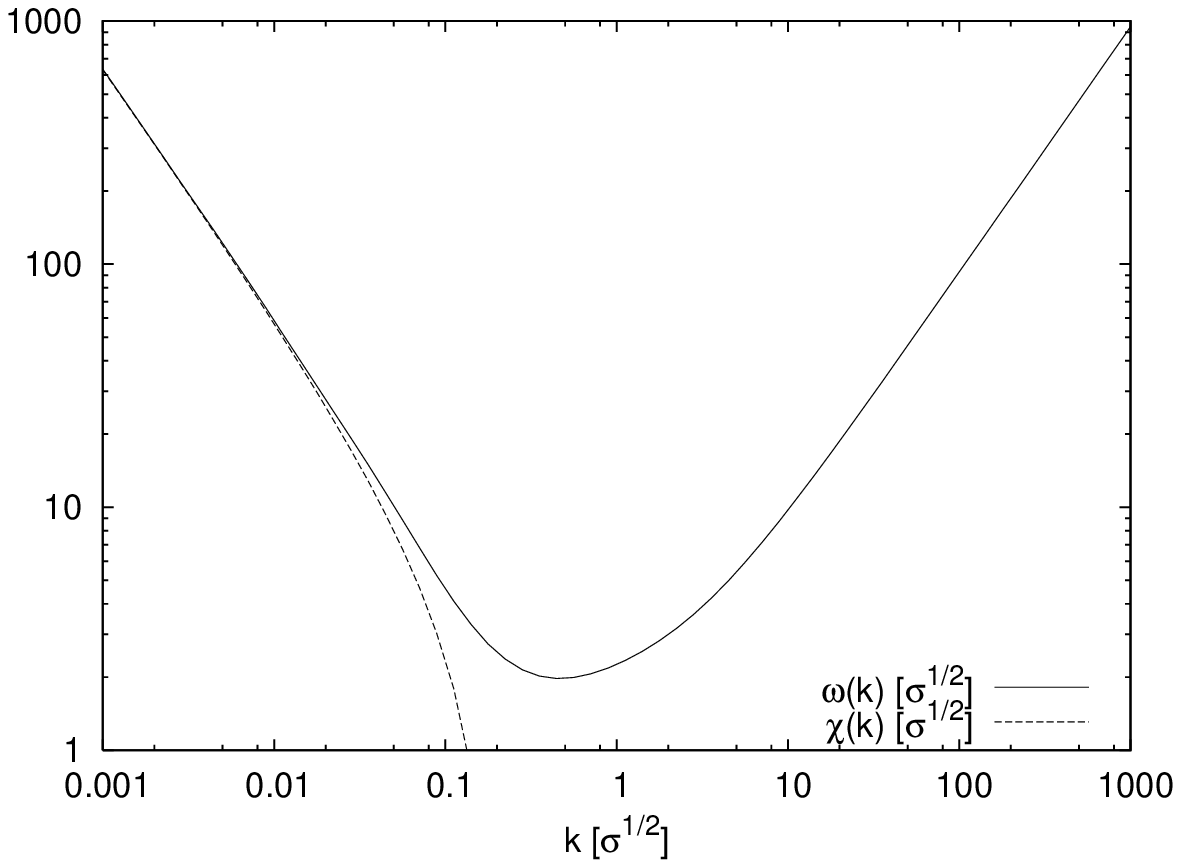}
  \caption{The ghost form factor $d (k)$ (left panel) and the gluon energy
$\omega (k)$ and the curvature $\chi (k)$ (right panel)}
\end{figure}
 The infrared exponent $\beta
= 1$ together with the sum rule (\ref{2}) implies an infrared linearly diverging
gluon energy, see fig.\ 1, which is a manifestation of confinement, and also
produces a linear rising static color potential, shown in fig.\ 2. 
\begin{figure}[ht]
  \includegraphics[height=.24\textheight]{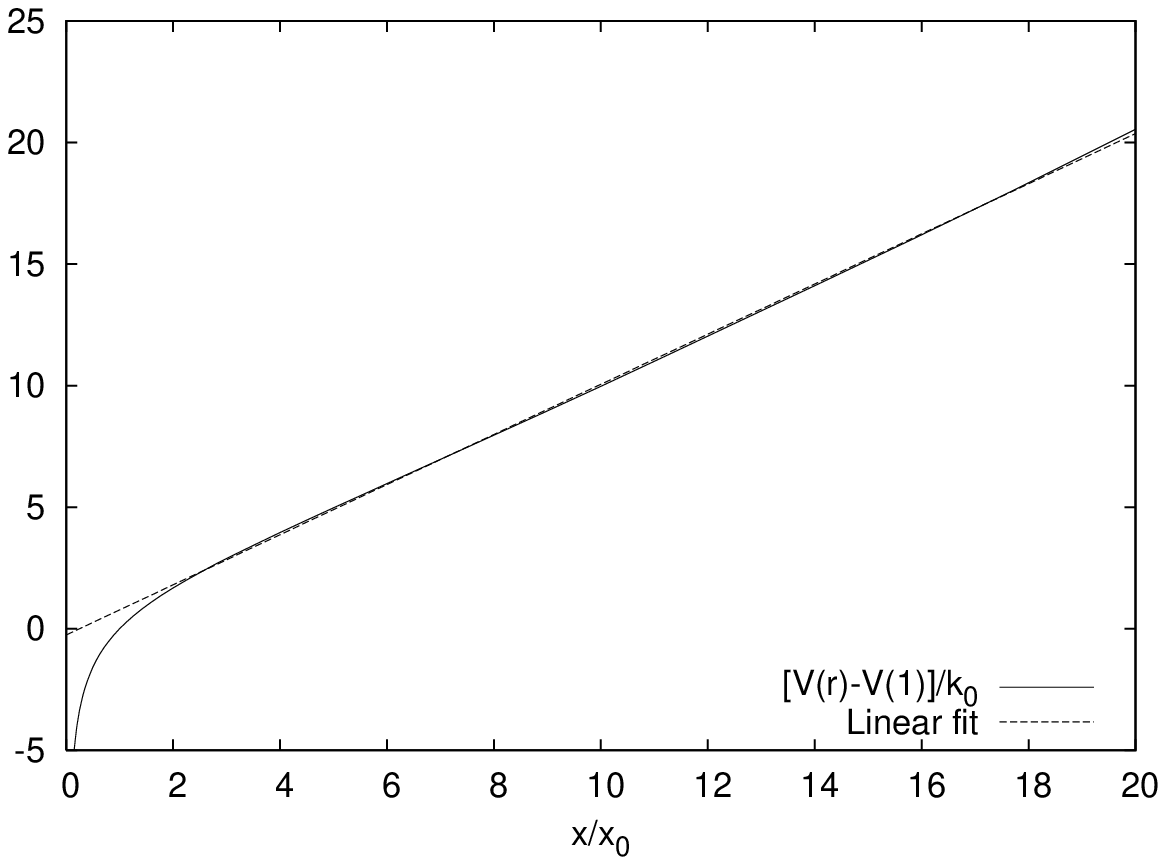}
  \includegraphics[height=.237\textheight,clip=]{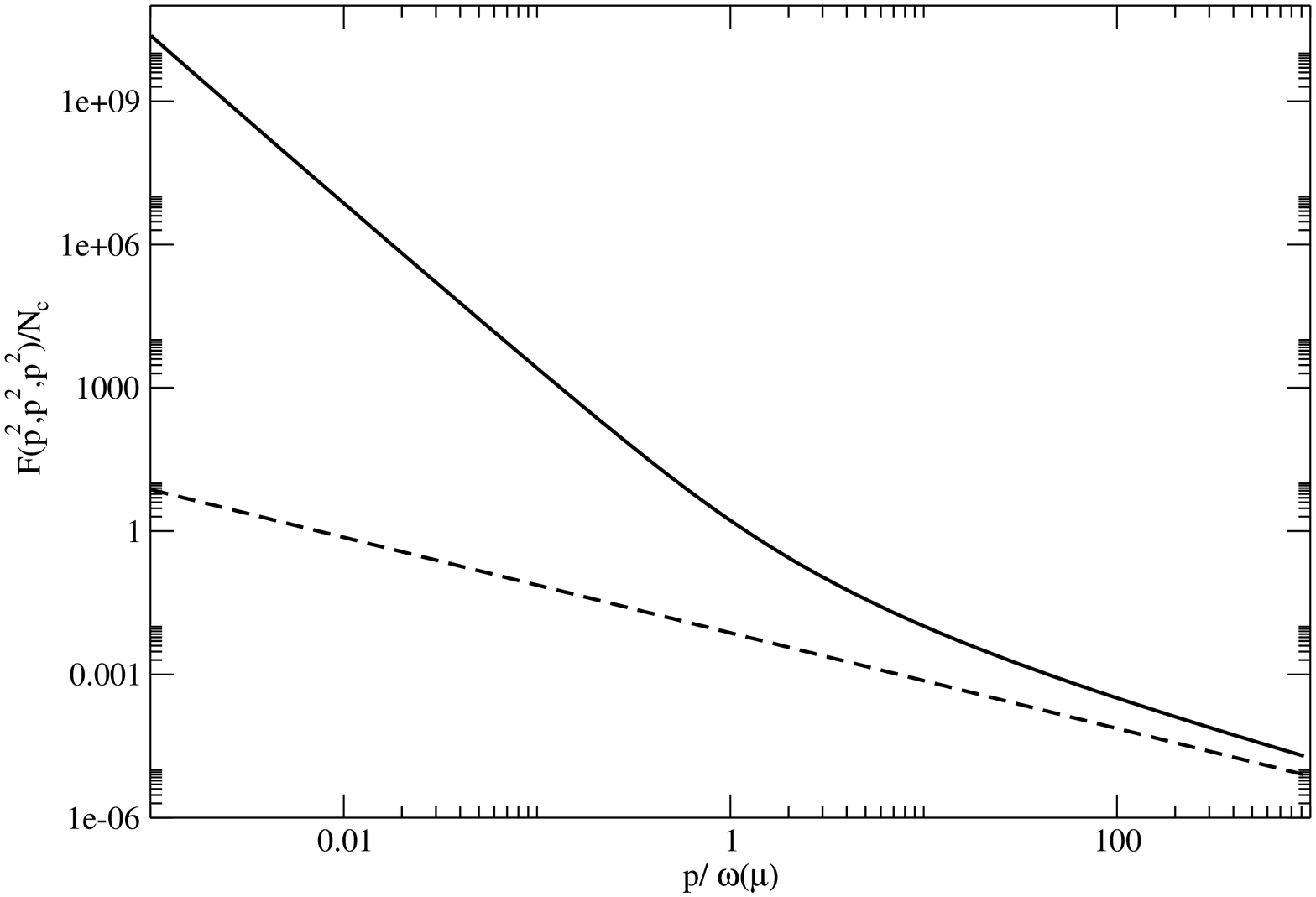}
  \caption{Left panel: The static Coulomb potential. Right panel: The three-gluon vertex at the symmetric point, full line with non-perturbative propagators, dashed line with perturbative propagators.}
\end{figure}
Let me stress that it is absolutely
crucial to keep fully the Faddeev-Popov determinant $J[A]$ in both the
Hamiltonian and the integration measure. The Faddeev-Popov determinant
converts the flat integration measure $\int \cD A$ into the Haar measure of the
gauge group \cite{Reinhardt:1996fs}. 
When the Faddeev-Popov determinant is neglected \cite{Szczepaniak:2001rg} the gluon energy becomes infrared finite and the linear rise in the static color potential is lost. On the other hand, the obtained
infrared behavior is quite robust with respect to changes of the variational
wave functional. Extending the variational ansatz, eq. (\ref{G2}), to arbitrary
$\alpha$ (where $\alpha = \frac{1}{2}$ and $\alpha = 0$ correspond to the wave
 functional used in refs. \cite{Feuchter:2004mk} and \cite{Szczepaniak:2001rg}, respectively), it can be shown that up to two 
 loops
 in the energy, the variational solution and in particular the ghost and gluon
 propagators are actually independent of $\alpha$, see ref.\ \cite{Reinhardt:2004mm}. Other quantities like the
 three-gluon vertex are, however, more sensitive than the energy and do depend 
 on
 $\alpha$. 

The leading contribution to the three-gluon vertex is given by the
 triangle diagram with an internal ghost loop.
In this order the three-gluon vertex has the same color structure as the bare
one and can be expressed by a single form factor
\be
\label{6}
\Gamma^{a b c}_{i j k} \lk p_1, p_2, p_3 \rk = f^{a b c} \lk i (p_1)_j 
\delta_{i
k} F  \lk p_1, p_2, p_3 \rk + \mbox{permutations} \rk \hk .
\ee
The form factor $F$ is shown in fig. 2 for the symmetric point $p_1 = p_2 =
p_3$. The infrared analysis performed in ref. \cite{Schleifenbaum:2006bq} shows $F (p^2) \sim
(p^2)^{\frac{d}{2} - 2 - \frac{3}{2} \beta}$. Inserting here $d = 3$ and $\beta
= 0.85$ from ref. \cite{Feuchter:2004mk}, one finds $F (p^2) \sim (p^2)^{-
1.775}$, which is in perfect agreement with the numerical result shown in fig.
2, from which one extracts $F (p^2) \sim (p^2)^{- 1.77}$. We have also
investigated the ghost-gluon vertex and found that, like in Landau gauge,  it becomes bare when one of the external ghost momenta vanishes,
so that its dressing can be ignored, at least in the infrared.

\no
We can define the RG-invariant running coupling
constant,
\be
\label{4XX}
\alpha (k) = \frac{16}{3} \frac{g^2_R}{4 \pi} p^5 G^2 (p) D (p) \, ,
\ee
from the ghost-gluon vertex \cite{Fischer:2005qe}, where $G(p)$ and $D(p)$ are the ghost and gluon propagators, respectively. Due to the sum rule (\ref{2}) for the infrared exponents this coupling has an
infrared fixed point which for the DSE solutions with $\beta = 1.0$ is given by
$\alpha (0) = 16 \pi/(3 N_c) \approx 16.76/N_c$, see ref.\ \cite{Schleifenbaum:2006bq}. This value obtained in the infrared analysis is in excellent agreement with the value of $\alpha(0)=16.94/N_c$, see \cite{Epple}, which is obtained from the numerical solution of the DSE presented in fig.\ 1. Let us also stress that this result disagrees with the value for the infrared fixed point obtained in the Coulomb gauge limit of the interpolating gauges \cite{Fischer:2005qe}.

\section{The 't Hooft loop}

The class (\ref{G2})  of wave functionals yield (up to two loops 
in the energy) the same Dyson-Schwinger equations independent of $\alpha$.  To test our wave functional and to
 get more detailed information on the structure of the
  Yang-Mills vacuum,  observables more sensitive than the
   energy density should be calculated. Moreover, the
  confining Coulomb potential obtained above does not
   necessarily guarantee that 
  our wave functional describes indeed a confining vacuum
   state. The reason is 
  that the Coulomb potential is only an upper-bound to the
   true static quark 
  potential \cite{Zwanziger}. Thus a confining Coulomb
   potential is a necessary but not a sufficient condition for a confining vacuum.

In pure Yang-Mills theory the order parameter of confinement is the temporal 
Wilson loop, which obeys an area law in the confined phase and a 
perimeter law in the 
deconfined phase. Unfortunately, this quantity is difficult to calculate in the 
continuum theory due to the path ordering. Fortunately, there exists another (dis-)order parameter of Yang-Mills 
theory which is easier to calculate in the continuum theory. This is the spatial 
 't Hooft loop which is dual to the temporal Wilson 
 loop in the sense that it obeys a perimeter law in the confined phase and an
  area law in the deconfined phase \cite{'tHooft:1977hy}.

The operator $V (C)$ of 't Hooft's disorder parameter is defined for a spatial 
loop $C$ by the following relation \cite{'tHooft:1977hy}:
\begin{eqnarray}
\label{defV}
V (C_1) W (C_2) = Z^{L (C_1, C_2)} W (C_2) V (C_1)\quad , \quad  W (C) = \frac{1}{N} tr \, P 
\exp (i \oint\limits_c A)
\end{eqnarray}
where $W (C)$ is the spatial Wilson loop, $Z$ is a (non-trivial) 
center element of the gauge group 
 ($Z = -1$ for SU(2)) and $L (C_1, C_2)$ denotes the Gaussian linking number
  between the two spatial loops $C_1$ and $C_2$. Unfortunately 't Hooft did not
   give an explicit realization of the operator $V (C)$ in the continuum theory,
    but defined the operator implicitly as having the effect of a singular (on 
    the loop $C$) gauge transformation.

The physical meaning of the 't Hooft loop defined by equation (\ref{defV}) is 
recognized by noticing that a time independent (spatial) center vortex 
${\mathcal A} (C_1)$ whose magnetic flux is localized on the closed loop $C_1$, 
produces 
the Wilson loop $W \big[ \mathcal{A} (C_1) \big] (C_2) = Z^{L (C_1, C_2)}
$. It follows that the 't Hooft loop operator $V (C)$ can be 
 interpreted as a center vortex creation operator. 
 The operator $V (C)$ which generates a center vortex gauge field 
 $\mathcal{A} (C)$ i.e., $ V (C) \Psi (A) = \Psi \big( A + \mathcal{A} (C) 
 \big)$,
is given by 
\begin{eqnarray}
\label{G6a}
V (C=\partial \Sigma) = \exp \left(i \int d^3 x \mathcal{A}^a_i (\Sigma, x) \Pi^a_i (x)\right).
\end{eqnarray}
In reference \cite{Reinhardt:2002mb} it was proven that the operator (\ref{G6a}) with the vortex 
field 
\begin{eqnarray}
\label{G7a}
\mathcal{A}^a_i (\Sigma, x) = \zeta^a \int_{\Sigma} d^2 \tilde{\sigma_i} 
\delta^3 (x - \bar{x}( \sigma))
\end{eqnarray}
with $\zeta = \zeta^a T_a$ being a co-weight vector defined by $\exp (i \zeta) 
= Z$ actually satisfies the defining relation (\ref{defV}). 
Since $\Pi^a_i (x)$ is the operator of the electric field, 
the 't Hooft loop measures the electric flux (projected on the co-weight 
$\zeta = \zeta^a T_a$) through the closed loop $C$, while
 the (spatial) Wilson loop $W (C)$  measures the magnetic flux through $C$.
The operator $V(C)$ is not manifestly gauge 
invariant, but produces a gauge 
 invariant result, when acting
  on gauge invariant states $\Psi (A^U) = \Psi (A)$.

The wave functional in Coulomb gauge satisfies Gauss' law (which ensures gauge 
invariance) and has, hence, to be considered as the restriction of the gauge 
invariant wave functional to transverse gauge fields. Therefore by implementing
the Coulomb gauge by the standard Faddeev-Popov method, we obtain 
for the 't Hooft loop 
\begin{eqnarray}
\label{G10}
\langle V (C) \rangle = \int \mathcal{D} A \, J (A) \, \Psi^{\ast} (A) \, \Psi 
(A + \mathcal{A}^\perp (C) \big) \hk ,
\end{eqnarray}
where $\mathcal{A}^\perp (C = \partial \Sigma)$ is the
 transverse part of the center vortex field $\mathcal{A} (\Sigma)$ (\ref{G7a})
  given by \cite{Reinhardt:2001kf}
\begin{eqnarray}
\label{G11}
\mathcal{A}^{\perp} (\partial \Sigma,x) = - \zeta \oint_{\partial 
\Sigma} d \tilde{\sigma}_{ik} \partial^{\bar{x}}_k \, \frac{1}{4\pi| x - \bar{x}|} 
 \quad , \quad d \tilde{\sigma}_{i k} = \Sigma_{i k \ell} \, d 
\bar{x}_{\ell}
\end{eqnarray}
which manifestly depends only on the loop $C = \partial \Sigma$ but not 
on the enclosed surface $\Sigma$ as the field (\ref{G7a}) does.
It is straightforward to calculate from eq. (\ref{G11}) 't Hooft's disorder
 parameter inserting for $\Psi (A)$ the wave functional determined previously \cite{Feuchter:2004mk}.
 After straightforward calculations, one finds 
\begin{eqnarray}
\label{G13}
\langle V (C) \rangle  = \exp (- S) \hk , \hk
S (C) = \int^{\infty}_0 d q \, K (q) \, h (C, q) \hk ,
\end{eqnarray}
where
\begin{eqnarray}
\label{G15}
K (q) = \frac{1}{2} \omega (q) \, \big( 1 - \frac{\chi^2 (q)}{\omega^2 (q)} \big)
\end{eqnarray}
contains the whole information about Yang-Mills vacuum while the quantity 
$h (C, q)$ is exclusively determined by the geometry of the considered loop.
 To simplify the calculations of this quantity, we consider a
  planar circular loop with radius $R$.  A detailed analysis shows that the $R$ dependence of $S (R)$
is determined by the infrared behavior of $K (q)$. Given that $\omega (q)$ 
and $\chi (q)$ are both infrared divergent and differ only by a finite constant,
one finds 
\begin{eqnarray}
\label{G19}
K (q \to 0) = \big( \omega (q) - \chi (q) \big)_{q \to 0} = c
\end{eqnarray}
where $c$ is a finite renormalization constant which can be chosen arbitrarily. 
Choosing $c = 0$, which implies that the infrared wave functional reduces 
to the strong coupling limit $\Psi [A] = 1$, one finds indeed a perimeter law 
${S} (R) = \tilde{\kappa} R$,
while for $c \neq 0$ one obtains ${S} R \sim R \, \ln( R / R_0)$.
Finally, if one ignores the curvature $(\chi = 0)$ in eq. (\ref{G15}), one finds an 
area law ${S} (R) = \tilde{\varrho} R^2$.
Thus again we find the curvature to be crucial for the confinement properties
 of the theory. \\[0.3cm]

\section{Summary and Conclusions}

In my talk I have discussed recent results obtained in the Hamilton approach 
to Yang-Mills theory in Coulomb gauge. Using ghost dominance, an infrared 
analysis of the Schwinger-Dyson equations has revealed both quark and gluon
 confinement. The gluon energy is infrared diverging implying the absence of 
 asymptotic gluon states from the physical spectrum, and the static Coulomb 
 potential is linearly rising. These results are confirmed by a full numerical
  solution of the Schwinger-Dyson equations. While the ghost-gluon vertex 
  remains basically unrenormalized and in particular becomes the bare one when
   the ghost momentum vanishes, the three-gluon vertex is strongly infrared 
   divergent.
As a first non-trivial test of the variationally determined vacuum wave
 functional, we have calculated 't Hooft's disorder parameter choosing 
 the renormalization constants of the gap equation in such a way that the 
 wave functional reproduces in the infrared the strong coupling limit. Then 
 the 't Hooft loop shows indeed a perimeter law, indicating a confining vacuum.

\begin{theacknowledgments}
Discussions with M. Quandt and P. Watson are gratefully acknowledged. This work was supported by DFG-Re856/6-1,2.
\end{theacknowledgments}



\bibliographystyle{aipprocl} 


\IfFileExists{\jobname.bbl}{}
 {\typeout{}
  \typeout{******************************************}
  \typeout{** Please run "bibtex \jobname" to optain}
  \typeout{** the bibliography and then re-run LaTeX}
  \typeout{** twice to fix the references!}
  \typeout{******************************************}
  \typeout{}
 }

\end{document}